\documentstyle[sprocl]{article}

\def\be{\begin{eqation}}
\def\ee{\end{equation}}
\def\bea{\begin{eqnarray}}
\def\eea{\end{eqnarray}}

\begin{document}

\title{Virtual Color Superconductivity--Status and Perspective
\footnote{Talk given at the ``Symposium on the Frontiers of Physics at
                  Millennium'', Beijing, China, Oct. 8-10, 1999.}
}
\author{S. Ying}
\address{Department of Physics, Fudan University, Shanghai 200433, China\\
         E-mail: sqying@fudan.edu.cn}

\maketitle
\begin{abstract}
   The status of our understanding of the properties and
manifestations of color superconductivity at zero and low density
is briefly reviewed.  Future possibility is highlighted.
\end{abstract}
\section{Introduction}

QCD at finite temperature and zero density condition is
relatively well understood compared the finite density
situation. The technical reasons are well known. Physically,
quantum decoherence increases as temperature gets high making the
behaviour of the system essentially classical. On the other hand,
density effects do not reduce quantum coherence. One has to deal
with the quantum effects. A brief review is given here about 
the current status of our understanding of 
the role played by the color superconducting phase in the strong
interaction ground state close to zero density 
and in hadron structure.

\section{High Density Limit}

The ground state of QCD at high enough density is most likely to
be in a color superconducting phase. This is because quarks on a
high Fermi surface carry high momentum and produce large momentum
transfer in scattering, so that the two quark interaction kernel
in color triplet channel can be approximated well by an
attractive one gluon exchange term.  The attractive interaction
generates the BCS instability that creates a gap of order $
\Delta_{BCS} \sim e^{-c/g} $ with $c$ a constant. The properties
of the color superconducting phase are actively studied in the
recent literature base on various approaches like the instanton
model, the one gluon exchange approximation improved by
renormalization group treatment and the contact 4-fermion
interaction model.

No matter what models are adopted for the discussion, the fact
that color superconducting phase should be the ground state of
the high density quark matter is quite clear according to well
established BCS theory for superconductivity in non-relativistic
condensed matter systems. A gap of order of a few 100 MeV is
guaranteed at not so high density since the natural scale of the
problem is of order 1 GeV and the coupling between quarks is of
order 1. Otherwise, fine tuning is required.
 
The general consensus on the properties of the color
superconducting phase of quark matter at asymptotically high
density and zero temperature is that the ground state is in a
color flavor locked state involving up, down and strange quarks.
As the density is lowered, the color flavor locked state ceases
to be the true ground state of the system, the true ground state
is in a color superconducting phase involves only the up and down
quarks. Such a color superconducting phase is again disfavoured as
the density is further lowered to some critical density
$\rho_c$. The color superconducting phase stops to manifest at a
macroscopic scale when $\rho\le \rho_c$. The current estimate of
$\rho_c$ is so high that the physical effects of the color
superconducting phase can only be possible to manifest in a
neutron star in the conventional understanding of the problem
based mainly on the physical picture gained in the study of
non-relativistic many body systems.

Since there are already reviews on this subject\cite{Shaf} at
high density, I shall mainly concentrate on a brief review of my
own efforts in the past 10 years (1989-1999), which actually
started the study of vacuum color superconductivity.  I however
looked at the problem with a different perspective.

\section{The Possible Role of the Color Superconducting Phase at
Low or Zero Density}

 What is the fate of the color superconducting phase at low
density then?  One thing is quite clear that the 
attraction between quark pairs in a color triplet state still
exist. It is evidenced by the existence of baryons which are made
up of three quarks. One of the reasons besides confinement for
the color superconducting phase to disappear at low density is
due to the competition from the quark--antiquark pairing.  Since
as the density is lowered, the antiquark excitation due to
interaction becomes less suppressed. A sufficient number of the
interaction generated antiquarks begin to pair with the quarks of
the system to condense in the ground state, which cause the well
know spontaneous breaking of the chiral symmetry.

 
In order to study the phase and the neighbourhood of the strong
interaction vacuum state, the present framework of finite density
theory based on non-relativistic space-time is found to be
inadequate.  Inconsistencies in the logic flow of physical
argumentation and unphysical solution occur when one goes beyond
the quasi-particle picture.  Thus not only the kinematics of
individual particles has to be relativistic, but also the
covariance related to the relativistic space-time transformation
for the whole system at the quantum field level has to be
implemented in a way that is consistent to quantum mechanics.

  The results of the efforts include: 1) the model building and
studying for vector\cite{plb,ann1} and scalar\cite{ccst1} diquark
condensation and many of the technical development\cite{ann1} 2)
the introduction of an 8--component real representation for
fermions\cite{plb,ann1,9611,ann2,tmu}, which is not identical to
the conventional 4-component representation at finite
density\cite{9611,tmu}. The advantage of using the 8--component
theory to study the vacuum properties is discussed. It also
enable me to isolate some of the differences near zero density
between my work and the later commonly adopted ones\cite{npb}
which contain a continuous mixing between the color
superconducting phase and the chiral symmetry breaking phase in
the parameter space where no metastable phase is found. The
difference originates from a different coupling between quarks in
the diquarks. While the ``reality condition'' adopted here
requires that quark and mirror quark\footnote{It is identical to
a quark at zero density.} are correlated to condense, other
approaches couple quark and mirror antiquark for the correlated
diquarks\cite{convs}, which can naturally be mixed with the
chiral symmetry breaking phase. Such a difference disappear at
high density.  3) the possibility that the effects of a
metastable phase can manifest in local observables and
propagators \cite{ann2,clst} 4) the development of a consistent
local finite density theory\cite{ann2,ep1,tmu}, which combines
all of the above considerations. The effects of statistical
blocking in the chiral symmetry breaking phase and the
spontaneous CP violation in the color superconducting phase was
discovered as a result\cite{ann2,ep1,tmu}. Also discussed is the
dynamics of the statistical gauge field in various phases.  It is
believed that the statistical blocking effects could provide a
mechanism\cite{ann2} to prevent the dissolution of a nucleon
inside a nucleus\cite{birs}, which is in direct contradiction to
the experimental fact that a nucleon inside a nucleus keeps most
of its identity.
 

Is what are talked about above correspond to reality? To answer it,
some of the empirical knowledge about a nucleon, which serves as a
carrier of the information about the virtual phase of the vacuum state,
are analyzed. 

The chiral properties of a nucleon is studied\cite{pcac1,pcac2}, to
search for a vector type of virtual color superconducting phase. There
are some hints.  More detailed works are needed. The recent discovery
of a large anapole moment for a nucleon\cite{SAMP} that can not be
accounted for by conventional electroweak theory\cite{Muslf} may also
be a hint for a vector type virtual color superconducting phase since
the order parameter here breaks parity which can induce an axial type
of coupling for a photon to a nucleon\cite{ann1}. The electromagnetic
properties of a nucleon is studied next base on the mechanism of
partial breaking of the electromagnetic gauge symmetry\cite{DIS} in a
color superconductor.  The Gerasimov-Drell-Hearn sum rule, which is
currently unable to be saturated by experimental data, is a possible
observable for both scalar and vector type of color superconducting
phase\cite{GDH}. A collection deep inelastic scattering data for a
lepton on a nucleon is analyzed, the results, especially the measured
polarized structure function for a nucleon at small Bjorken x,
indicates the existence of a virtual superconducting
phase\cite{DIS}. Other evidences, which is less model independent, is
also considered\cite{ccst1}. In the near future, at least two possibile
developements could be foreseen: 1) the nucleon $\Sigma$--term problem,
which is made worse by the analysis \cite{an1,an2} of most recent data
on pion nucleon scattering, could be resolved
\cite{pcac2} by using the local theory developed here \cite{Sig} 2)
type of the CP/T violation data obtained in the KTeV Collaboration
should be analyzed\cite{PTRV} in more details to search\cite{SW} for
possible anomalies.

\section*{Acknowledgements}

  This work was support by National Natural Science Foundation of China.
I would like to thank Prof. A. W. Thomas and the CSSM of Univeristy of 
Adelaide for hospitality during which this write up is done.

\end{document}